\documentclass{article}

\PassOptionsToPackage{numbers, compress}{natbib}
\usepackage[final]{neurips_2024}
\usepackage{subcaption}

\usepackage[utf8]{inputenc} 
\usepackage[T1]{fontenc}    
\usepackage{hyperref}       
\usepackage{url}            
\usepackage{booktabs}       
\usepackage{amsfonts}       
\usepackage{nicefrac}       
\usepackage{microtype}      
\usepackage{xcolor}         

\usepackage{graphicx}
\usepackage{comment}

\usepackage{booktabs}
\usepackage{graphicx}

\title{Graph Neural Networks for Efficient AC Power Flow Prediction in Power Grids}

\author{
Amir Talebi \\
Electrical and Computer Engineering\\
North Carolina State University\\
Raleigh, NC, USA \\
\texttt{amirtalebi879@gmail.com} \\
\And
Kaixiong Zhou \\
Electrical and Computer Engineering\\
North Carolina State University\\
Raleigh, NC, USA \\
\texttt{kzhou22@ncsu.edu} \\
}

\begin{document}

\maketitle

\begin{abstract}
This paper proposes a novel approach using Graph Neural Networks (GNNs) to solve the AC Power Flow problem in power grids. AC OPF is essential for minimizing generation costs while meeting the operational constraints of the grid. Traditional solvers struggle with scalability, especially in large systems with renewable energy sources. Our approach models the power grid as a graph, where buses are nodes and transmission lines are edges. We explore different GNN architectures, including GCN, GAT, SAGEConv, and GraphConv to predict AC power flow solutions efficiently. Our experiments on IEEE test systems show that GNNs can accurately predict power flow solutions and scale to larger systems, outperforming traditional solvers in terms of computation time. This work highlights the potential of GNNs for real-time power grid management, with future plans to apply the model to even larger grid systems.

\end{abstract}

\section{Introduction}

The Optimal Power Flow (OPF) problem is fundamental and one of the most important optimization in power system operations \cite{cain2012}. It involved the minimization of generation costs while meeting load demands and satisfying the various grid constraints such as generator capacities (Active and reactive power limitsy), voltage magnitude and phase limits and line thermal limits. Efficient and accurate solutions to the OPF problem are essential for real-time power system management, especially as modern power systems become increasingly complex with the integration of variable renewable energy sources and increased of the number of nodes.

Traditional OPF solvers, such as Newton-Raphson or Interior Point Methods (IPOPT), often face challenges in scalability and computational efficiency. As the number of nodes (buses) and edges (transmission lines) in the network increases, the computational complexities grow significantly and make real-time applications difficult. These methods also tend to be computationally intensive, limiting their use in large-scale grids or in systems with high variability introduced by renewable energy sources. The exact formulation of the problem is commonly referred to as ACOPF and the approximation as DCOPF. Incorporating AC dynamics results in a non-convex nature of OPF problem due to nonlinearities in the power flow equations. Authors in \cite{talebi2022} proposed a Mixed-Integer Linear Programming (MILP) approach to improve the resilience of distribution systems. Their approach demonstrates how optimization can be leveraged to address non-linearities and system constraints, such as those found in the AC Power Flow (ACOPF) problem. This foundational work provides useful insights for the integration of advanced optimization techniques with power system management.

Recent advancements in Machine learning and deep learning, motivated by the possibility of accurately generating
large amounts of data, and particularly Graph Neural Networks, offer a promising solutions to address these challenges. GNNs are designed to handle graph-structured data effectively which makes them well-suited for modeling power grids where buses can be represented as nodes and transmission lines as edges. By leveraging GNNs, we aim to develop a model that can predict OPF solutions efficiently, maintaining high accuracy while significantly reducing computation time.

In this paper, we propose a novel approach using GNNs to improve the accuracy and scalability of OPF predictions. We will explore multiple GNN topologies, such as multi-hop message-passing GNNs, which allow for deeper propagation of information across the grid, and attention-based GNNs that assign varying importance to different buses in the network based on their influence on OPF solutions. The model will be trained on standard IEEE test cases, and its performance will be compared against traditional solvers, including Newton-Raphson. The objective is to develop a model that scales well to large power grids and is capable of real-time grid management.











\section{Related Work}
\label{gen_inst}

\begin{figure}[htbp]
  \centering
  \includegraphics[width=0.5\textwidth]{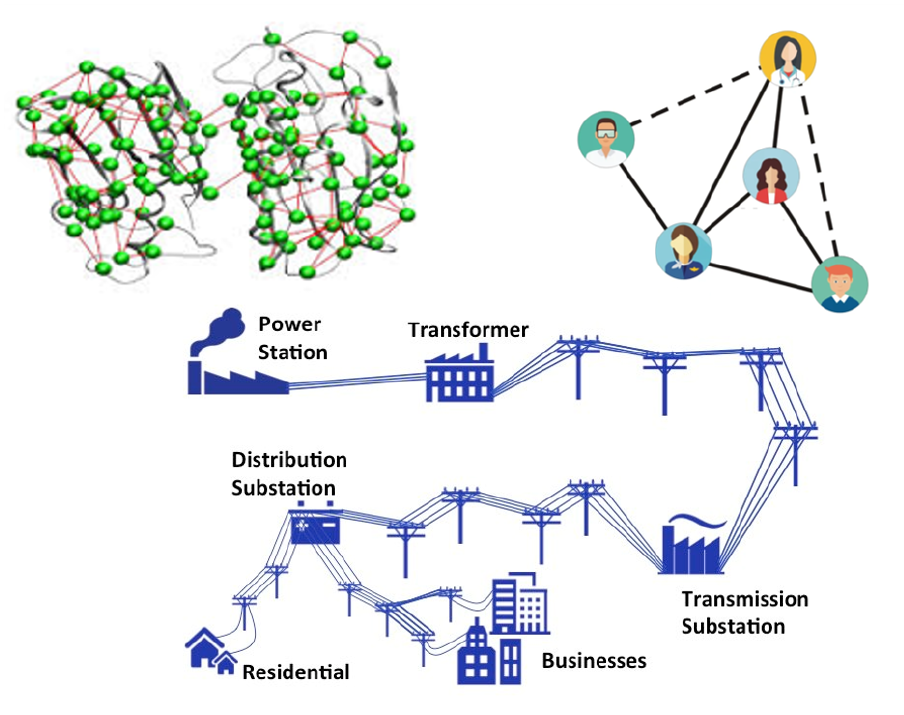}
  \caption{Proteins \cite{vijayabaskar2011}, social networks \cite{ahmad2020}, and electrical power grids \cite{hosein2016} are all graphs}
  \label{fig:sample-figure}
\end{figure}

In recent years, machine learning (ML) approaches have been introduced to approximate or solve OPF more efficiently. Among these, multi-layer perceptrons (MLPs) were some of the earliest ML models applied to power system problems. Studies such as \cite{guha2019} applied fully connected network (MLP) to imitate the output of ACOPF and predict power flow variables and generator outputs, However, MLPs face significant challenges with local minima and overfitting, especially when applied to large-scale power systems with non-linearities. They also struggle to model the graph structure of the power grid, where buses (nodes) and transmission lines (edges) are interconnected in a highly structured manner. The failure to leverage this topological information leads to inaccurate results and poor performance, especially in generalization across different grid topologies.

To address these shortcomings, Graph Neural Networks (GNNs) have emerged as a more suitable architecture for power system applications. GNNs excel in handling graph-structured data which make them suitable for power grids, where buses and transmission lines naturally form a graph. GNNs use a message-passing mechanism, allowing each node to aggregate information from neighboring nodes, which mirrors the flow of electrical voltages and currents between interconnected buses in the power system.

Authors in \cite{chen2020} were among the first to apply GNNs to the OPF problem. They developed a GNN-based model that uses imitation learning to predict OPF solutions based on the outputs of traditional solvers like IPOPT. The model demonstrated significant improvements in computation time compared to traditional solvers while maintaining high accuracy, particularly for the IEEE-30 and IEEE-118 bus systems. This approach successfully exploits the grid topology to improve efficiency and provides a foundation for future research in using GNNs for OPF.

\cite{donti2021} proposed a Topology-Aware Graph Neural Network that incorporates both the spatial structure of the grid and physical constraints into the learning process. Their model introduces AC-feasibility regularization and ensures that the GNN’s predictions commit to the physical laws of power flow, Kirchhoff’s laws, and generator limits. This method enhanced the generalizability of GNN models across different grid topologies and more importantly ensured that solutions were physically feasible, a critical requirement in real-world power systems.

Further extending the capabilities of GNNs, \cite{lopez2023} introduced a Typed Graph Neural Network (TGNN) approach in their work on power flow analysis. TGNNs differentiate between different types of buses (e.g., generator buses, load buses) in the power grid, allowing the model to treat each node type according to its operational role. This added complexity in node classification improves the accuracy of power flow predictions, especially in cases with diverse bus types and heterogeneous grid configurations. Their experiments with typed GNNs showed better performance compared to standard GNN models, particularly in grids with mixed bus types.

Authors in \cite{liao2022} provide a comprehensive review of the application of GNNs in power systems, highlighting their advantages in terms of scalability, generalizability, and performance when applied to complex, non-linear power system problems like OPF. The authors emphasize the ability of GNNs to handle dynamic grid configurations and temporal variations which make them effective in environments with high renewable penetration and fluctuating load conditions. Their review also covers various GNN architectures, including ChebNet and spectral-based GCNs, which have been applied to solve power flow and fault detection problems. The study’s key contribution is its focus on task analysis, where GNNs outperform conventional deep learning models by leveraging the intrinsic graph structure of the power grid. They also discuss the critical challenges of data availability and the need for advanced regularization techniques to ensure physical feasibility in GNN predictions.

\cite{yang2020} introduced a model that combines Proximal Policy Optimization (PPO) with GNNs for OPF. Their work leverages the decision-making capabilities of PPO to control generator outputs in a power grid while using GNNs to model the spatial relationships within the grid. Their experiments showed that this approach outperforms traditional solvers like DCOPF in both cost minimization and constraint satisfaction, particularly in dynamic environments where grid topology or load conditions change.

Additionally, \cite{donon2020} explored the application of Graph Neural Solvers (GNS) to directly solve power flow equations by minimizing violations of Kirchhoff’s laws. Their work focused on developing a graph-based solver that could scale efficiently with grid size while ensuring commitment to physical laws, laying the groundwork for GNN-based models that solve power flow equations directly without the need for traditional optimization solvers.

More recently, \cite{fan2021} introduced a comparative study between Koopman Operator-based models and GNNs for learning power grid dynamics. This work highlighted the advantages of GNNs in capturing the spatio-temporal dynamics of power grids, demonstrating their utility not only for static optimization problems like OPF but also for transient analysis and fault detection. The study showed that GNNs could outperform Koopman models in terms of generalizability and adaptability to changing grid conditions.

Incorporating more recent work, \cite{jeddi2021} proposed a Graph Attention Network (GAT) that integrates physical constraints such as Kirchhoff's laws. This hybrid model focuses on incorporating domain-specific physics into the learning process, ensuring both scalability and physical accuracy. The GAT model allows for efficient propagation of information across nodes (buses), addressing scalability concerns while preserving physical feasibility. This makes it highly suitable for real-time OPF applications in complex, renewable-integrated grids.

\cite{wu2022} demonstrate the use of GNNs in a probabilistic context. Their focus is on using attention mechanisms to prioritize critical nodes in a grid, especially under uncertain conditions caused by renewable energy sources. This marks a shift from deterministic to probabilistic approaches in power flow and addresses the challenges posed by stochastic variations in renewable energy generation.

The progression from traditional solvers and MLPs to GNNs represents a shift towards more scalable and efficient OPF solutions. The key advantage of GNNs lies in their ability to incorporate the grid topology which improves both prediction accuracy and computational efficiency. However, challenges remain, particularly in ensuring that GNNs can handle real-time grid operations and adapt to dynamic topologies in a computationally feasible manner.

\section{Methodology}
\label{methodology}

\subsection{AC Power Flow Problem Formulation}

AC Power Flow analysis aims to determine the voltage magnitudes (\(V\)) and phase angles (\(\delta\)) at each bus in a power grid while ensuring active (\(P\)) and reactive (\(Q\)) power balance across the network. The nonlinear power flow equations are given by:

\[
P_i = V_i \sum_{j=1}^{n} V_j (G_{ij} \cos(\delta_i - \delta_j) + B_{ij} \sin(\delta_i - \delta_j)),
\]
\[
Q_i = V_i \sum_{j=1}^{n} V_j (G_{ij} \sin(\delta_i - \delta_j) - B_{ij} \cos(\delta_i - \delta_j)),
\]

Where \(P_i\) and \(Q_i\) represent the net real and reactive power injections at bus \(i\), \(V_i\) and \(\delta_i\) are the voltage magnitude and angle, and \(G_{ij}\) and \(B_{ij}\) are the real and imaginary components of the admittance matrix.

The admittance matrix components \(G_{ij}\) (conductance) and \(B_{ij}\) (susceptance) play a crucial role in influencing the AC power flow results. They represent the electrical properties of the transmission lines connecting buses, and their values directly affect the power flow between buses. \(G_{ij}\) controls the active power loss and \(B_{ij}\) governs the power transfer through reactive power loss. These interactions between buses mirror the message-passing mechanism in Graph Neural Networks (GNNs), where each node aggregates and updates information from its connected neighbors.

In our approach, voltage magnitude \(V\) and phase angle \(\delta\) are updated dynamically as node features during training, which forms the basis for the AC Optimal Power Flow solution. By modeling the power flow equations through GNNs, we effectively propagate and learn critical interactions between buses. Additionally, bus types (PV, PQ, Slack) were incorporated as node features, capturing the specific operational roles of each bus type and enabling the GNN to learn how these roles influence the overall power flow dynamics.

\subsection{Dataset Generation}

The datasets for our experiments were generated using the standard IEEE test systems with realistic \%40 variations in load demands. The main objective was to create diverse scenarios that effectively evaluate the robustness and generalization capability of our GNN models. To achieve this, we used pandapower, a Python-based library tailored for power system analysis, and leveraged its built-in IEEE test cases and power flow capabilities.
The load variations were designed to simulate both daily and seasonal patterns and incorporate realistic fluctuations in load values. This network includes predefined bus configurations, loads, generators, and transmission lines which is an ideal test case for load flow analysis. The script generates 10 separate datasets, each containing 10000 samples, which results in a total of 100,000 data points per bus system configuration as shown in Figure ~\ref{tab:dataset_structure}

\begin{itemize}
    \item \textbf{IEEE 14-Bus System:} A small and simple transmission network containing 14 buses, 5 generators (including the slack bus), 11 loads, and 20 branches. It’s widely used for educational purposes and serves as a good starting point for testing basic load flow algorithms. Its smaller size allows for quick evaluation of model capabilities before scaling to larger systems.

    \item \textbf{IEEE 30-Bus System:} A medium-sized network with 30 buses, 6 generators, 21 loads, and 41 branches. It is commonly used for testing load flow and OPF algorithms. Its diverse generation and load configurations help assess the scalability and generalization of GNN models, simulating realistic operational conditions with larger variations in load and generator setpoints.

    \item \textbf{IEEE 57-Bus System:} A more complex transmission network containing 57 buses, 7 generators, 42 loads, and 80 branches. It is commonly used for load flow analysis, contingency analysis, and testing the robustness of power flow solvers. It poses a challenging scenario due to its larger network size and increased interconnections, requiring the GNN model to effectively propagate information across a wider spatial area. It helps evaluate the model’s ability to handle medium to large-sized systems and test its robustness in predicting power flow variables accurately under diverse operating conditions.

    \item \textbf{IEEE 118-Bus System:} A large and realistic transmission network with 118 buses, 19 generators, 99 loads, and 186 branches. It is widely used for studies involving OPF, contingency analysis, and voltage stability. It provides a comprehensive test for scalability, allowing us to validate the generalization capability of our model on real-world, large-scale power systems.
    
\end{itemize}

\begin{figure}[htbp]
    \centering
    \begin{minipage}{0.45\textwidth}
        \centering
        \includegraphics[width=\textwidth]{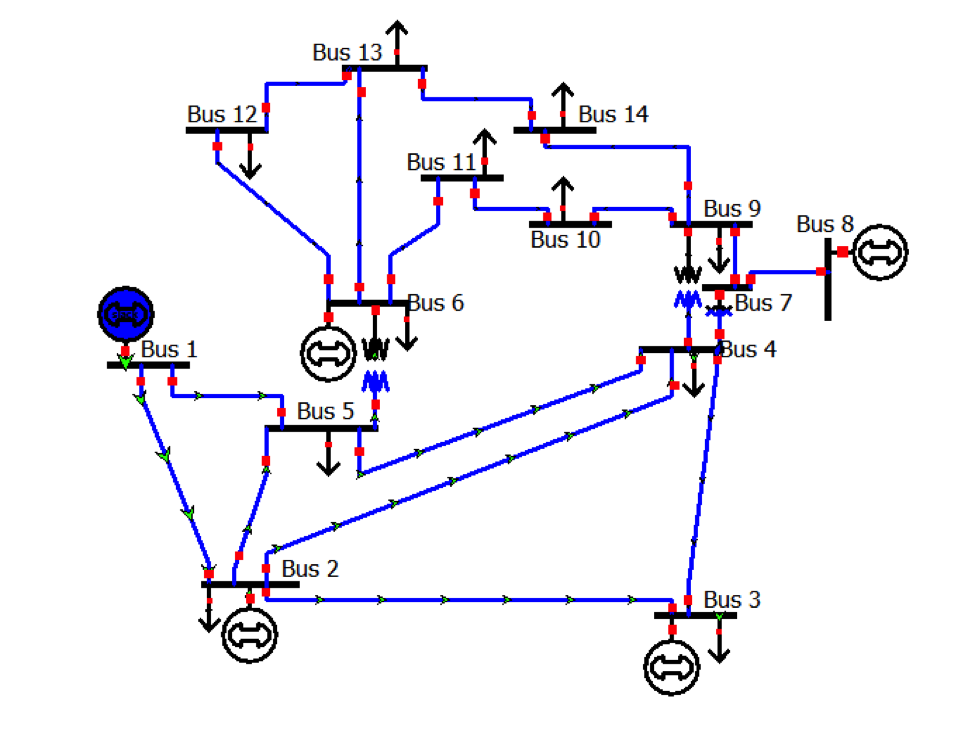}
        \caption*{\textbf{(a)} IEEE 14-Bus System}
        \label{fig:ieee14}
    \end{minipage}
    \hfill
    \begin{minipage}{0.45\textwidth}
        \centering
        \includegraphics[width=\textwidth]{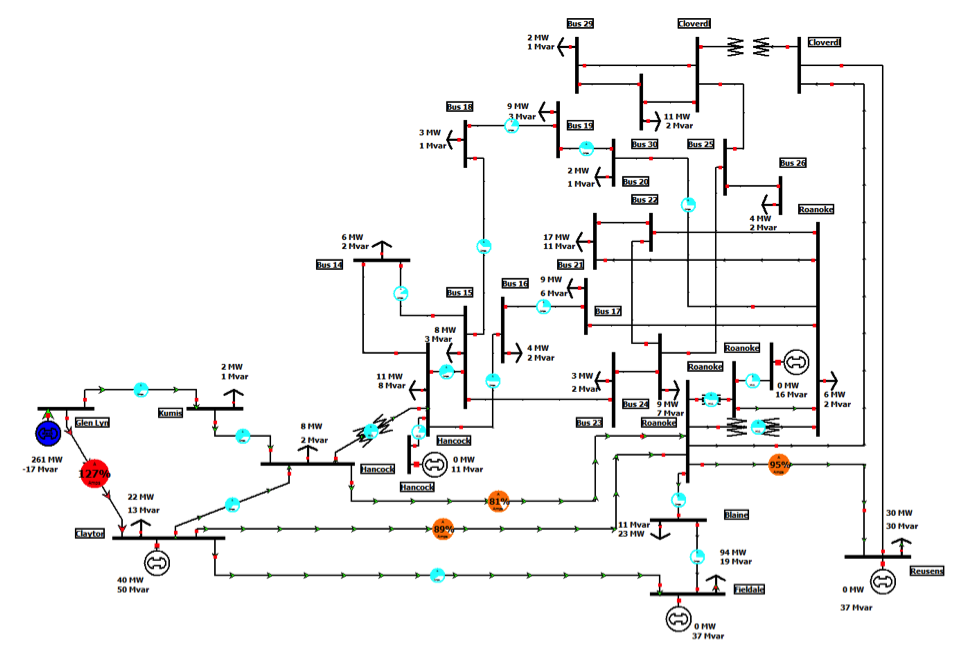}
        \caption*{\textbf{(b)} IEEE 30-Bus System}
        \label{fig:ieee30}
    \end{minipage}
    
    
    \begin{minipage}{0.45\textwidth}
        \centering
        \includegraphics[width=\textwidth]{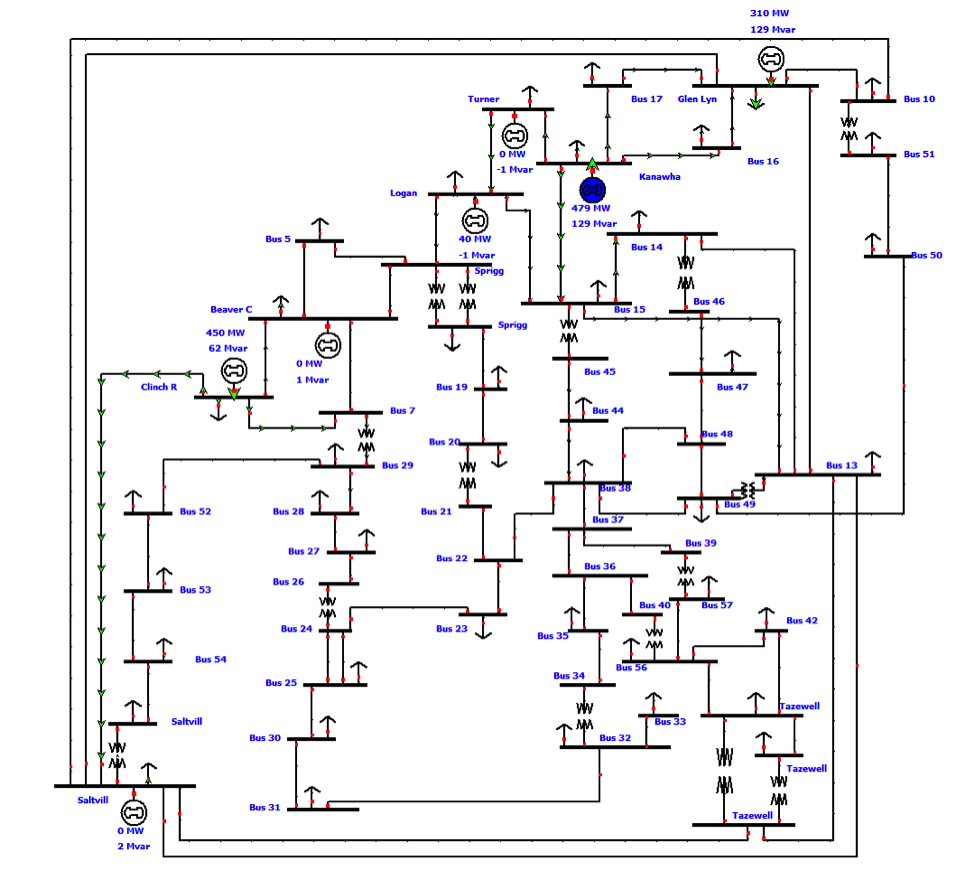}
        \caption*{\textbf{(c)} IEEE 57-Bus System}
        \label{fig:ieee57}
    \end{minipage}
    \hfill
    \begin{minipage}{0.45\textwidth}
        \centering
        \includegraphics[width=\textwidth]{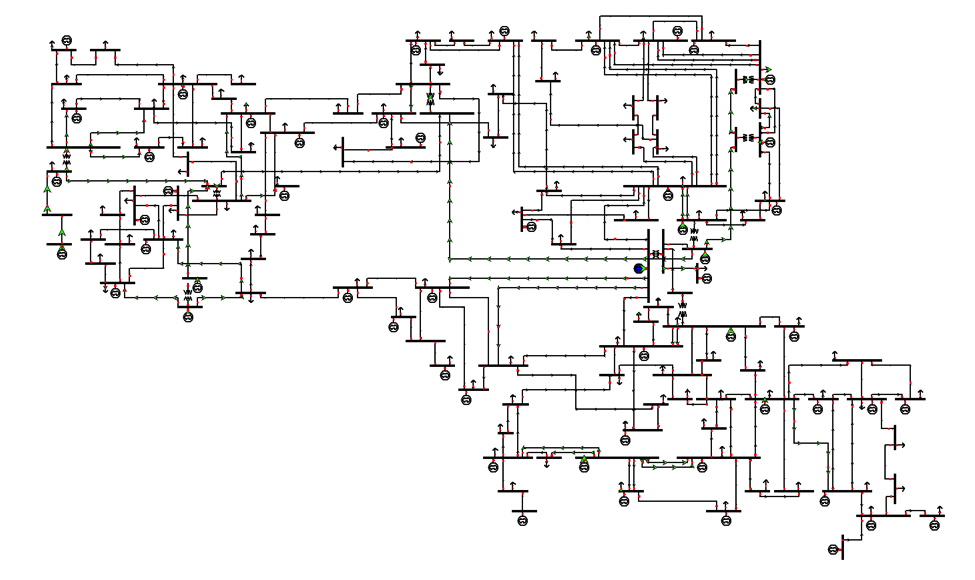}
        \caption*{\textbf{(d)} IEEE 118-Bus System}
        \label{fig:ieee118}
    \end{minipage}
    
    \caption{Diagrams of IEEE Bus Systems: 14-Bus, 30-Bus, 57-Bus, and 118-Bus configurations.}
    \label{fig:ieee_bus_systems}
\end{figure}

\begin{table}[htbp]
    \centering
    \caption{Dataset Structure for IEEE 30-Bus System}
    \label{tab:dataset_structure}
    \begin{tabular}{|c|c|c|c|}
        \hline
        \textbf{Data Point} & \textbf{Node} & \textbf{Input Features (P, Q, V, $\delta$)} & \textbf{Output Features (V, $\delta$)} \\
        \hline
        {Data Point 1} & 1  & $P_1$, $Q_1$, $V_1$, $\delta_1$ & $V_1$, $\delta_1$ \\
                                      & 2  & $P_2$, $Q_2$, $V_2$, $\delta_2$ & $V_2$, $\delta_2$ \\
                                      & 3  & $P_3$, $Q_3$, $V_3$, $\delta_3$ & $V_3$, $\delta_3$ \\
                                      & 4  & $P_4$, $Q_4$, $V_4$, $\delta_4$ & $V_4$, $\delta_4$ \\
                                      & ... & ... & ... \\
                                      & 30 & $P_{30}$, $Q_{30}$, $V_{30}$, $\delta_{30}$ & $V_{30}$, $\delta_{30}$ \\
        \hline
        {Data Point 10,000} & 1  & $P_1$, $Q_1$, $V_1$, $\delta_1$ & $V_1$, $\delta_1$ \\
                                           & 2  & $P_2$, $Q_2$, $V_2$, $\delta_2$ & $V_2$, $\delta_2$ \\
                                           & 3  & $P_3$, $Q_3$, $V_3$, $\delta_3$ & $V_3$, $\delta_3$ \\
                                           & 4  & $P_4$, $Q_4$, $V_4$, $\delta_4$ & $V_4$, $\delta_4$ \\
                                           & ... & ... & ... \\
                                           & 30 & $P_{30}$, $Q_{30}$, $V_{30}$, $\delta_{30}$ & $V_{30}$, $\delta_{30}$ \\
        \hline
    \end{tabular}
\end{table}

\newpage

Daily load variation is influenced by typical human activity and industrial operations, creating distinct patterns throughout the day. During the Morning Ramp-Up (6 AM - 9 AM), load demand increases as residential and commercial activities begin, reaching 60 to 70\% of the base load. The Midday Peak (10 AM - 3 PM) sees a load increase due to industrial and commercial energy usage, peaking at 110 to 120\% of the base load. The Evening Peak (5 PM - 9 PM) results in another higher peak as people return home, significantly increasing residential load, again reaching 110 to 120\% of the base load. Finally, during the Nighttime Drop (11 PM - 5 AM), load decreases significantly during late-night hours, dropping to 60 to 70\% of the base load.

Seasonal variations also affect load due to heating and cooling needs, influenced by weather and climate conditions. In the Winter Peak, loads are higher due to increased heating demand in colder climates, reaching 1.2 to 1.4 times the base load. The Summer Peak occurs with higher loads due to increased air conditioning usage in hot climates, reaching 1.1 to 1.3 times the base load. During the Spring/Fall Load, the load is moderate due to milder weather conditions.

These \%40 realistic variations ensure that our models are tested under different operating conditions, capturing both high-demand scenarios and low-demand scenarios. This helps in evaluating the model’s ability to generalize and maintain performance across a wide range of load conditions.

\subsection{Ground Truth Generation}

Ground truth data was generated using the Newton-Raphson method to solve the AC Power Flow equations for each load scenario. The Newton-Raphson method was chosen due to its robustness and fast convergence, making it ideal for solving non-linear power flow equations accurately. We also briefly considered Gauss-Seidel and Fast-Decoupled methods for comparison but found Newton-Raphson to be superior in accuracy and convergence rate.

\subsection{GNN Architecture and Topologies}

The following topologies will be tested to compare the effectiveness of different GNN architectures in solving the AC power flow problem. Each topology offers distinct strengths, such as handling local versus long-range relationships, incorporating attention mechanisms, and scaling to larger grid sizes:

\subsubsection{Graph Convolutional Networks \texttt{GCN}}
GCN is a basic form of GNN where each node aggregates features from its immediate neighbors. Its simplicity makes it a strong first model to set a benchmark. GCN is widely used for node classification and regression tasks and serves as a baseline model. It can handle local relationships but may struggle with capturing long-range dependencies in the graph.

\subsubsection{Graph Attention Networks \texttt{GATConv}} 
GAT assigns different attention weights to each neighboring node during the aggregation process, helping the network focus on the most important nodes. This dynamic weighting of neighbors is particularly useful in OPF tasks when certain nodes (such as generator buses) have a greater influence on the power flow solution. The attention mechanism allows the model to prioritize critical nodes, potentially improving prediction accuracy.

\subsubsection {GraphSAGE (Sample and Aggregate) \texttt{SAGEConv}} 
GraphSAGE is designed for inductive learning and allows the model to sample a fixed-size neighborhood of nodes, making it scalable to large graphs. GraphSAGE is well-suited for AC power flow tasks in large power grids, as it can efficiently handle expanding network sizes. It aggregates sampled neighbor features which make it robust for generalizing to unseen nodes like slack bus which will be far away from some nodes.

\subsubsection {Graph Convolution \texttt{GraphConv}} 
GraphConv enhances feature learning by incorporating self-loops, where each node aggregates its feature along with its neighbors. GraphConv is useful in scenarios where enhanced node feature aggregation (like generator buses) can lead to better model performance. It is expected to perform well in grids where local node information, like the generation results of the PV buses are crucial.


The selected GNN topologies serve different purposes in our experiments:

\begin{itemize}
    \item \textbf{Baseline:} \texttt{GCN} serves as the benchmark model to establish a performance baseline for comparison with more complex architectures.
    
    \item \textbf{Attention-Based:} \texttt{GATConv} is used to evaluate whether incorporating attention mechanisms helps improve performance in OPF tasks, particularly by prioritizing influential nodes.
        
    \item \textbf{Scalability Test:} \texttt{SAGEConv} is included to assess the model’s scalability and performance on larger bus systems, leveraging its inductive learning capabilities.
    
    \item \textbf{Enhanced Node Feature Learning:} \texttt{GraphConv} is used to investigate whether incorporating self-loops and enhanced node feature aggregation improves the model's performance, especially for nodes like generator buses.
    
   \end{itemize}


This feature set ensures that the model can differentiate between bus types, which is crucial for accurate power flow predictions and was not explicitly handled in previous baseline models.

\section{Experimental Setup and Dataset Generation}

\subsection{Experimental Setup detail}

To evaluate the performance of our GNN models, we conducted a comprehensive experimental setup involving dataset generation, normalization, model training, and testing across multiple IEEE test cases (14-bus, 30-bus, 57-bus, and 118-bus systems). The details of each step are described below:

\subsubsection{Dataset Preparation}

We used pre-processed datasets generated from power flow simulations based on IEEE test systems (14-bus, 30-bus, 57-bus, and 118-bus). The datasets were loaded from Excel files and divided into training, validation, and test sets as follows:

\begin{itemize}
    \item \textbf{Train Dataset:} Consisted of 100\% of the first dataset file for each bus system.
    \item \textbf{Validation Dataset:} Consisted of 20\% of a separate second dataset file to ensure an unbiased validation process.
    \item \textbf{Test Dataset:} For each bus system, we used 20\% of 10 separate test datasets, evaluating the model on diverse load scenarios.
\end{itemize}

Each bus system contains 10 different scenarios, with 10,000 data points per scenario. The datasets include 40\% load variations for each scenario. The summary of the datasets for each bus system is shown in Table~\ref{tab:datasets}:.

\begin{table}[htbp]
    \centering
    \caption{Summary of Datasets}
    \label{tab:datasets}
    \begin{tabular}{lccc}
        \toprule
        \textbf{Dataset Type} & \textbf{Number of Scenarios} & \textbf{Number of Data Points} & \textbf{Load Variations} \\
        \midrule
        Train        & 10  & 10,000 & 40\% \\
        Validation   & 1   & 2,000  & 40\% \\
        Test         & 10  & 10,000 & 40\% \\
        \bottomrule
    \end{tabular}
\end{table}

\subsubsection{Feature Extraction and Dataset Creation}

The input features for each bus included the Active Power (\(P\)), Reactive Power (\(Q\)), Voltage Magnitude (\(V\)), and Voltage Angle (\(\delta\)). Also, the Bus Type (Slack, PV, PQ), is represented as one-hot encoded features.
We defined the bus type in Table~\ref{tab:bus_type_definitions}.

The target outputs were the voltage magnitude (\(V\)) and voltage angle (\(\delta\)) for each bus.

\begin{table}[htbp]
    \centering
    \caption{Bus Type Definitions and Known/Unknown Variables}
    \label{tab:bus_type_definitions}
    \begin{tabular}{lcccc}
        \toprule
        \textbf{Bus Type} & \textbf{Known Variables} & \textbf{Unknown Variables} \\
        \midrule
        \textbf{Slack Bus} & \(V\), \(\delta\) & \(P\), \(Q\) \\
        \textbf{PV Bus (Generator Bus)} & \(P\), \(V\) & \(Q\), \(\delta\) \\
        \textbf{PQ Bus (Load Bus)} & \(P\), \(Q\) & \(V\), \(\delta\) \\
        \bottomrule
    \end{tabular}
\end{table}

\subsubsection{Data Normalization}

The features and targets were normalized to improve model training stability. We computed the mean and standard deviation for both input features and targets. All features except the bus type were normalized using z-score normalization. The bus type was kept as categorical and not normalized. The denormalization process was applied to predictions during evaluation to obtain actual voltage magnitudes and angles.

\subsubsection{Graph Construction and Data Loaders}

The bus systems were represented as graph-structured data using PyTorch Geometric. The edge index was constructed from the \texttt{from\_bus} and \texttt{to\_bus} connections in the network topology. Each graph data object consisted of Node Features, Edge Index and Target Outputs. Edge Index Represents bidirectional connections between buses based on the transmission line data.
We used PyTorch DataLoaders for batching the graph data with a batch size of 16 for both training and validation.

\subsubsection{Training Procedure and Hyperparameter Tuning}

The GNN models were trained using the setup provided in Table~\ref{tab:hyperparameters}

\begin{itemize}
    \item \textbf{Optimizer:} Adam optimizer with an initial learning rate of \(5 \times 10^{-5}\) and L2 regularization (\(\lambda = 1 \times 10^{-6}\)).
    \item \textbf{Learning Rate Scheduler:} We used a ReduceLROnPlateau scheduler to adjust the learning rate based on validation loss improvements.
    \item \textbf{Batch Size:} A batch size of 16 was used for both training and validation phases. Increasing the batch size helped in convergence and noise avoidance. 
    \item \textbf{Epochs:} Models were trained for a maximum of 800 epochs with early stopping based on validation loss, using a patience of 100 epochs.
    \item \textbf{Dropout and Batch Normalization:} We used a dropout rate of 0.2 and enabled batch normalization to prevent overfitting and improve model generalization.
\end{itemize}

\begin{table}[htbp]
  \caption{Summary of Hyperparameters for GNN Models}
  \label{tab:hyperparameters}
  \centering
  \resizebox{\textwidth}{!}{
  \begin{tabular}{lll}
    \toprule
    \multicolumn{2}{c}{Hyperparameter Settings} \\
    \cmidrule(r){1-2}
    \textbf{Hyperparameter} & \textbf{Value} \\
    \midrule
    Learning Rate & $5 \times 10^{-5}$ \\
    L2 Regularization (Lambda) & $1 \times 10^{-6}$ \\
    Dropout Rate & 0.2 \\
    GNN Type & GCN / GAT / SAGEConv \\
    Input Features & 7 (\(V\), \(\delta\), \(P\), \(Q\), \(Bus Type (PV, PQ, Slack))\) \\    
    Number of GNN Layers & 2 \\
    GNN Layer 1 Size & 12 \\
    GNN Layer 2 Size & 12 \\
    Hidden Layer Size (Fully Connected) & 128 \\
    Output Size & $2 \times n\_bus$ (Voltage Magnitude and Angle for Each Bus) \\
    Batch Size & 16 \\
    Optimizer & Adam \\
    Learning Rate Scheduler & Exponential Decay (Decay Factor: 0.9 every 10 epochs) \\
    Total Trainable Parameters & Varies by GNN Type \\
    Early Stopping Patience & 20 epochs \\
    \bottomrule
  \end{tabular}
  }
\end{table}

\subsubsection{Model Evaluation and Testing}

The performance of the GNN models was assessed using the following metrics:
\begin{itemize}
    \item \textbf{Mean Squared Error (MSE):} Measures the average squared difference between predicted and true values of voltage magnitude (\(V\)) and phase angle (\(\delta\)).
    \item \textbf{Normalized Root Mean Squared Error (NRMSE):} Provides a normalized measure of error, allowing for comparison across different test cases.
    \item \textbf{\(R^2\) Score:} Indicates the proportion of variance in the target variables explained by the model, with higher scores indicating better predictive performance.
\end{itemize}

For each test dataset (10 different scenarios for each bus system), we loaded the data and prepared it for evaluation. The best-performing model (based on validation loss) was used for testing. Predictions were denormalized, and regression metrics (MSE, RMSE, NRMSE, MAE, \(R^2\)) were calculated. The test losses were saved and summarized across all datasets. This method helped us have a comprehensive assessment of model performance.

\subsection{Results and Preliminary Analysis}

We plotted the training and validation loss curves for the IEEE 14-bus, 30-bus, 57-bus, and 118-bus systems to visualize the model's convergence. The loss curves demonstrate consistent convergence across all test cases, indicating effective learning and generalization. Additionally, we compared the performance of different GNN architectures (GCN, GAT, and SAGEConv and GraphConv) using three evaluation metrics: NRMSE, \( R^2 \) score, and average training loss.

The models achieved consistent performance across all test cases, with NRMSE values below \(0.05\). The \( R^2 \) scores for all the models were consistently high and close to 1, which indicate a strong fit to the target data as shown in Figure. \ref{fig:r2_comparison}. Also it is noted that as the number of buses increase the training loss generally decrease. Increasing the batch size and reducing the dropout rate led to improved convergence and lower test loss. 

\begin{itemize}
    \item \textbf{GCN:}
    GCN showed higher NRMSE values across all bus configurations, particularly for the 30-bus and 57-bus systems. This indicates its challenges in capturing complex and long-range dependencies. Average test loss for GCN was higher which imply slower convergence.

    \item \textbf{GAT:}
    GAT demonstrated significant improvements in NRMSE compared to simple baseline, especially for the 30-bus and 57-bus systems which leverage its attention mechanism to focus on critical nodes (e.g., generator buses). This shows its efficiency in learning meaningful representations and a balanced results across between more complex architectures.

    \item \textbf{SAGEConv:}
    SAGEConv achieved the lowest NRMSE values across almost all bus configurations, demonstrating its scalability and superior performance in handling larger systems like the 118-bus network. Despite slightly higher test loss compared to GAT and GraphConv, SAGEConv showed strong generalization, as evidenced by its lower NRMSE across all configurations as shown in Figure. \ref{fig:nrmse_comparison}.

\begin{figure}[htbp]
    \centering
    \includegraphics[width=0.8\textwidth]{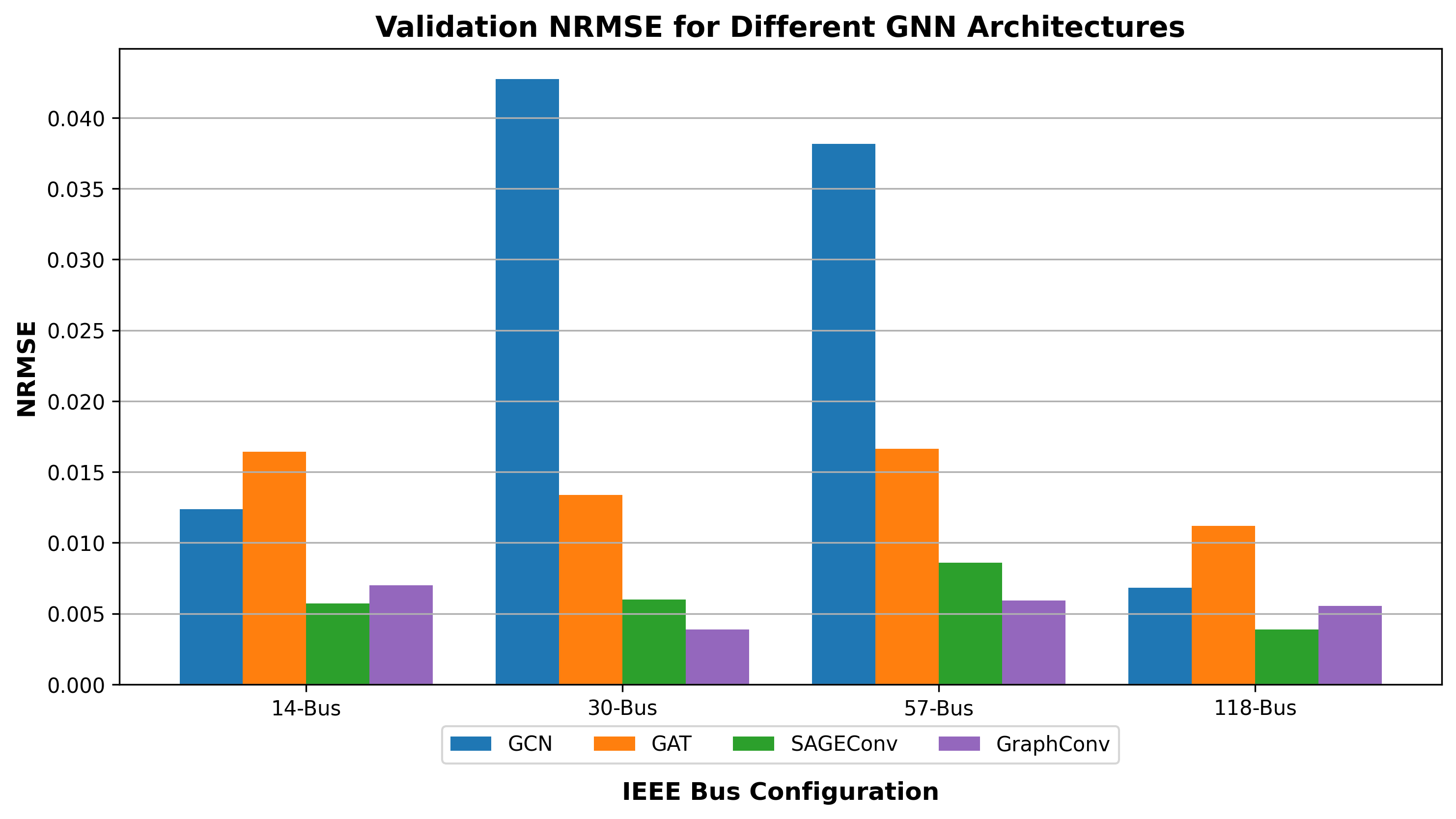}
    \caption{Comparison of NRMSE across different IEEE bus configurations using three GNN architectures: GCN, GAT, SAGEConv and GraphConv.}
    \label{fig:nrmse_comparison}
\end{figure}

\begin{figure}[htbp]
    \centering
    \includegraphics[width=0.8\textwidth]{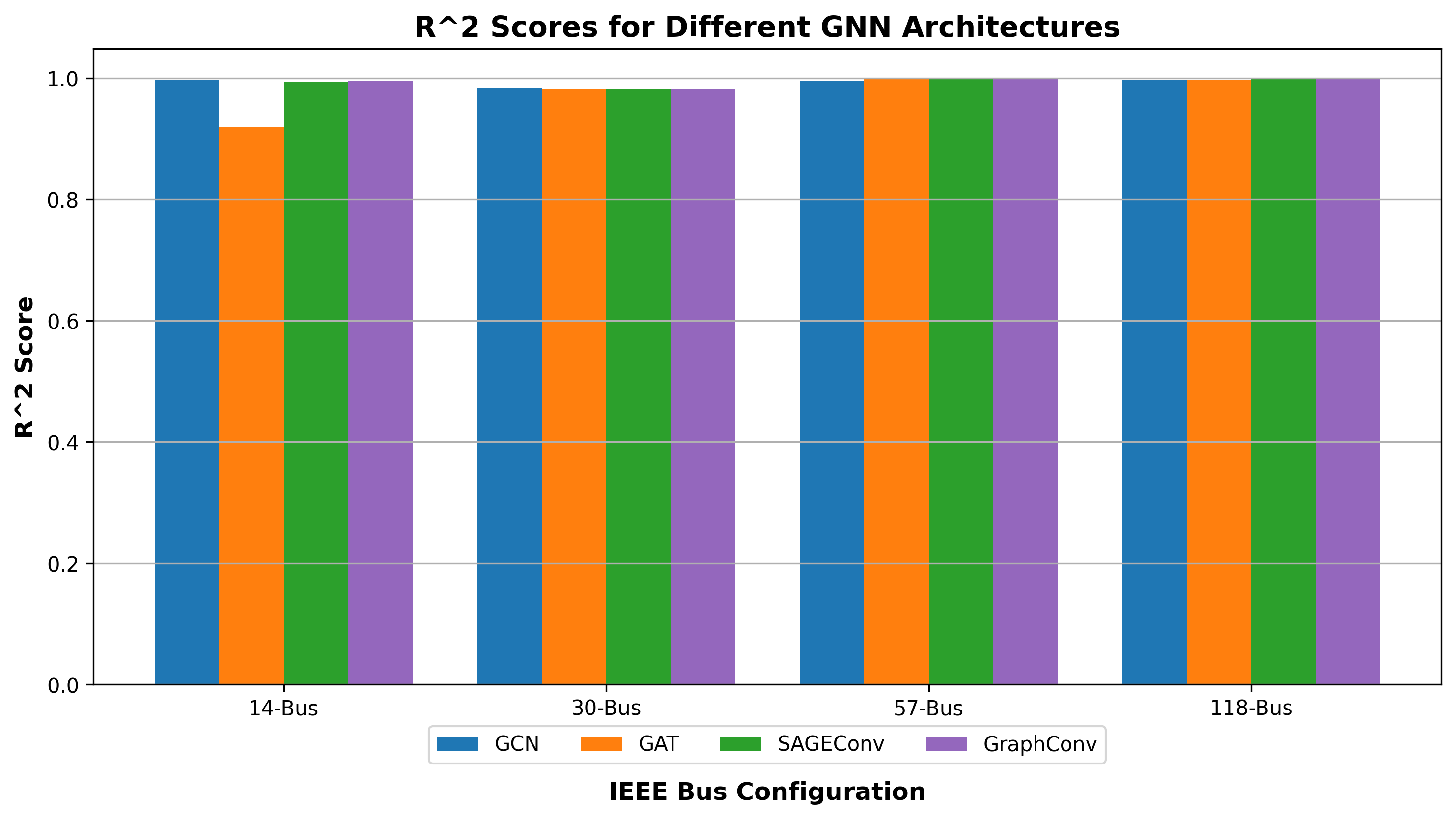}
    \caption{Comparison of \( R^2 \) scores for different IEEE bus configurations using GCN, GAT, SAGEConv, and GraphConv.}
    \label{fig:r2_comparison}
\end{figure}

\begin{figure}[htbp]
    \centering
    \includegraphics[width=0.8\textwidth]{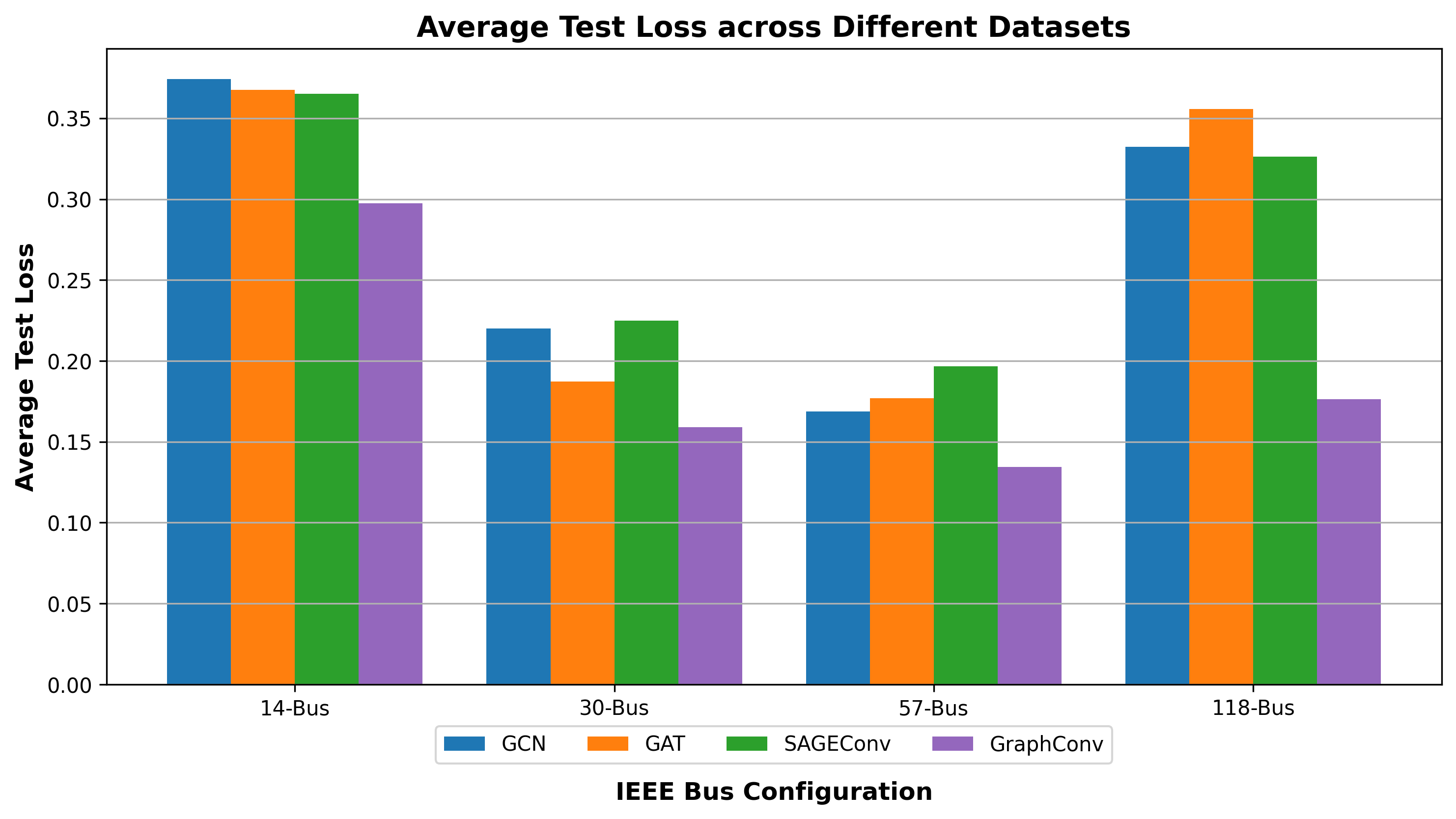}
    \caption{Comparison of average test loss across different IEEE bus configurations using GCN, GAT, SAGEConv, and GraphConv.}
    \label{fig:average_test_loss}
\end{figure}

\begin{figure}[htbp]
    \centering
    \begin{subfigure}{0.45\textwidth}
        \centering
        \includegraphics[width=\textwidth]{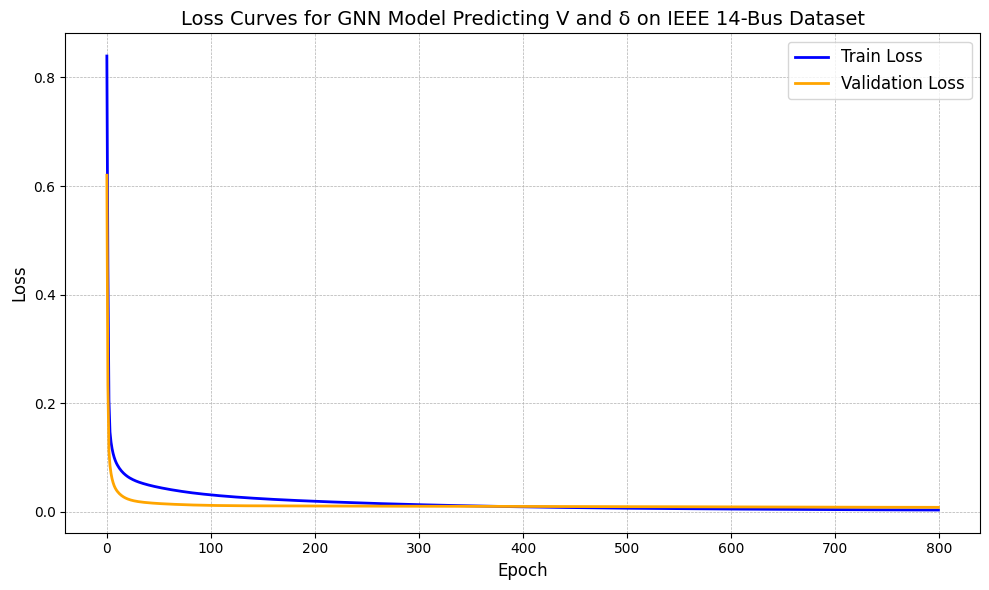}
        \caption{IEEE 14-Bus System}
    \end{subfigure}
    \hfill
    \begin{subfigure}{0.45\textwidth}
        \centering
        \includegraphics[width=\textwidth]{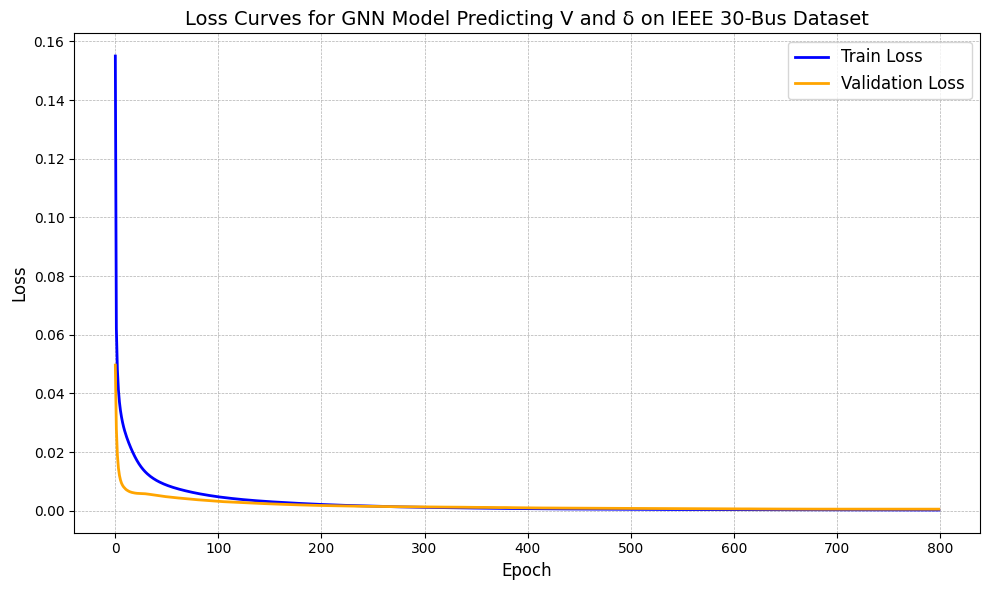}
        \caption{IEEE 30-Bus System}
    \end{subfigure}
    

    \begin{subfigure}{0.45\textwidth}
        \centering
        \includegraphics[width=\textwidth]{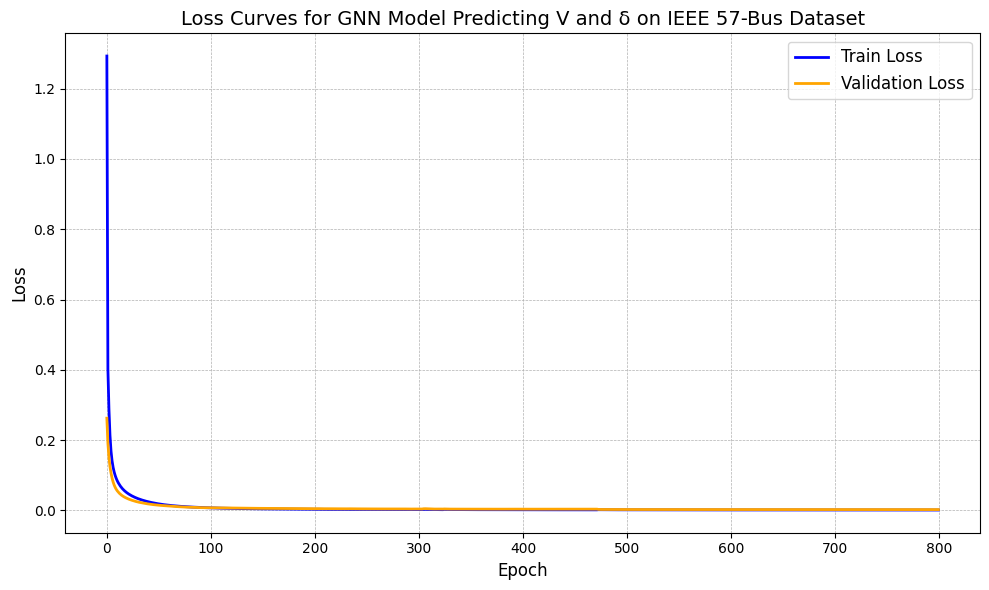}
        \caption{IEEE 57-Bus System}
    \end{subfigure}
    \hfill
    \begin{subfigure}{0.45\textwidth}
        \centering
        \includegraphics[width=\textwidth]{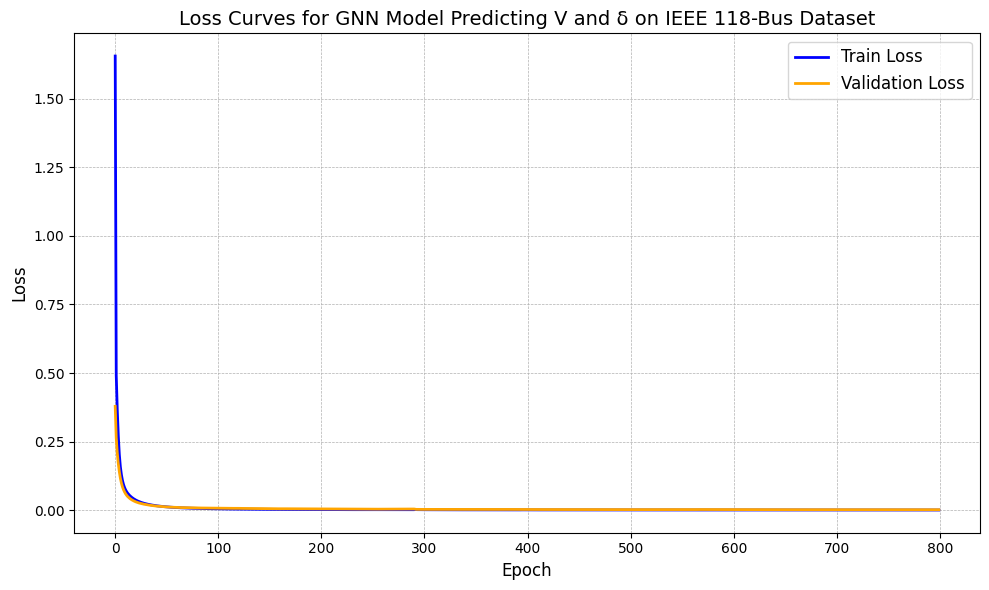}
        \caption{IEEE 118-Bus System}
    \end{subfigure}
    
    \caption{Training and validation loss curves for the IEEE 14-bus, 30-bus, 57-bus, and 118-bus systems using GCN Architecture.}
    \label{fig:loss_curves}
\end{figure}

    \item \textbf{GraphConv:}
    GraphConv along with SAGEConv consistently achieved the lowest NRMSE values across all bus configurations, This was expected as in AC power flow, the \(V\), \(\delta\), \(P\), and \(Q\) of each node will affect its AC power flow solution. This results in the lowest test loss across all network configuration as shown in Figure. \ref{fig:average_test_loss}.
    
\end{itemize}

\newpage

\section{Conclusion}

This paper presents a novel GNN-based approach for solving the AC Power Flow problem, incorporating voltage magnitude, voltage angle updates, and bus type as features. Preliminary results on IEEE test cases demonstrate strong performance. The consistent trend between training and validation losses suggests effective learning and good generalization.

The results highlight the strengths and weaknesses of each GNN architecture across various bus configurations. The findings suggest that attention-based models like GAT and scalable architectures like SAGEConv and GraphConv, which effectively handle large datasets, are promising choices for power flow prediction tasks. SAGEConv’s capacity to capture long-range dependencies, combined with GAT’s ability to prioritize influential nodes, as well as GraphConv with its effective handling of node-specific information through enhanced feature aggregation, highlights their potential for real-world applications in power systems.

The full implementation of the proposed model is available on GitHub: \href{https://github.com/Amirtalebi83/GNN-OptimalPowerFlow}{Amir Talebi GitHub Repository: GNN-OptimalPowerFlow}. \cite{Talebi2023}

\section{Future Work}

In future work, we plan to extend our analysis to large-scale systems to further assess the scalability of GNN-based models. Test systems such as the ACTIVSg 2000-Bus and Polish System Test Cases (e.g., 2383-bus and 9241-bus systems) are ideal for testing the scalability of GNN capability in solving AC power flow in massive grids.

We also aim to investigate the role of generator buses (PV and slack buses) in AC power flow problems. These buses play a key role in grid stability, and GNNs like SAGEConv, which capture long-range dependencies, may naturally prioritize them during training. Our future research will focus on understanding how GNNs focus on generator buses and whether SAGEConv’s message-passing mechanism enhances performance for predicting power flow variables. We also plan to analyze the learned node embeddings to check if generator buses receive more attention during training.

\clearpage

\bibliographystyle{unsrtnat}
\bibliography{neurips_2024}

\end{document}